\documentclass[numbers,webpdf,imaiai]{ima-authoring-template}

\usepackage[utf8]{inputenc}
\usepackage[T1]{fontenc}
\usepackage{amssymb}
\usepackage{mathtools}
\usepackage{dynkin-diagrams}
\usepackage{xcolor}

\newcommand{\Z}{\mathbb Z}
\newcommand{\Sp}{\mbox{Sp}}
\newcommand{\GL}{\mbox{GL}}
\newcommand{\F}{\mathbb F}
\newcommand{\Fq}{\F_q}
\newcommand{\Rnn}{\mathbb R ^n}
\newcommand{\Orb}{\mbox{Orb}}
\newcommand{\IG}{\mbox{IG}}
\newcommand{\inv}{^{-1}}

\newcommand{\tH}{{\widetilde H}}
\newcommand{\flis}[1]{\mathcal{#1}^{\rm iso}}
\newcommand{\HSp}{H_L}
\newcommand{\Hq}{\tH_2}
\newcommand{\SO}{\mbox{SO}}

\theoremstyle{thmstyletwo}
\newtheorem{proposition}{Proposition}
\newtheorem{lemma}{Lemma}
\theoremstyle{thmstylethree}
\newtheorem{remark}{Remark}

\numberwithin{equation}{section}

\begin{document}
\DOI{}
\copyrightyear{2026}
\vol{00}
\pubyear{2026}
\access{Advance Access Publication Date: Day Month Year}
\appnotes{Paper}
%\copyrightstatement{Published by Oxford University Press on behalf of the Institute of Mathematics and its Applications. All rights reserved.}
\firstpage{1}

\title[Entropies associated with orbits of finite groups]{Entropies associated with orbits of finite groups}

\author{Ryan Leal
\address{\orgdiv{Department of Mathematics}, \orgname{California Institute of Technology}, \orgaddress{\street{1200 E. California Blvd., Pasadena}, \postcode{91125}, \state{CA}, \country{USA}}}}
\author{Jingtong Sun
\address{\orgdiv{Department of Mathematics}, \orgname{California Institute of Technology}, \orgaddress{\street{1200 E. California Blvd., Pasadena}, \postcode{91125}, \state{CA}, \country{USA}}}}
\author{Juan Pablo Vigneaux*
\address{\orgdiv{Department of Mathematics}, \orgname{California Institute of Technology}, \orgaddress{\street{1200 E. California Blvd., Pasadena}, \postcode{91125}, \state{CA}, \country{USA}}}
\address{\orgdiv{Department of Linguistics}, \orgname{Northwestern University}, \orgaddress{\street{Evanston}, \postcode{60208}, \state{IL}, \country{USA}}}}

\authormark{R. Leal, J. Sun and J. P. Vigneaux}

\corresp[*]{Corresponding author: \href{mailto:jpvigneaux@northwestern.edu}{jpvigneaux@northwestern.edu}}

%\received{Date}{0}{Year}
%\revised{Date}{0}{Year}
%\accepted{Date}{0}{Year}

\abstract{For many groups, parabolic subgroups arise as stabilizers of flags of sets or vector spaces; the quotients by these subgroups parametrize orbits of flags, and  entropies appear as rates of exponential or superexponential growth of their cardinalities. The multiplicative ``chain rules'' relating these cardinalities induce, asymptotically, additive analogues for the entropies. Traditional formulas in information theory correspond to quotients of symmetric groups, a particular kind of reflection group: orbit cardinalities are multinomial coefficients, related to Shannon entropy. Quotients of general linear groups over a finite field can be treated similarly, with $q$-multinomials in place of multinomials and the Tsallis 2-entropy in place of Shannon's. Here we consider other finite reflection groups and classical groups over finite fields (split groups of Lie type). In both settings the groups are classified by Dynkin diagrams into infinite series $A_n$, $B_n$, $C_n$, $D_n$ and finitely many exceptional ones; the $A_n$ series consists of the symmetric groups (reflection case) and general linear groups (Lie case). Our treatment is uniform, based on classical invariant theory: every finite reflection group $W$ has well-defined \emph{degrees}, and the product of their $q$-deformations is the Poincar\'e polynomial of $W$ (Solomon's formula). Quotients of Poincar\'e polynomials count at once the cosets of parabolic subgroups of $W$ (at $q=1$) and the points of the flag varieties $G/P$ of the corresponding algebraic groups over $\mathbb F_q$ (at a prime power $q$). The series $B_n$, $C_n$, and $D_n$, studied here from an information-theoretic perspective for the first time, are linked to new entropic functionals: a \emph{reflective entropy} governing the reflection groups, and a \emph{Lie entropy} governing the associated groups of Lie type; each follows a distinctive chain rule. }

\maketitle

\section{Introduction}

\subsection{Motivation}\label{sec:intro_background}

A primary aim of this work is to connect symmetry in the sense of group actions and entropy in information theory. Consider the uniform probability distribution $\mathbb U : \omega \mapsto 1/|\Omega|$ on a finite set $\Omega$. This maximizes the entropy and also has the highest symmetry under permutations, but under arbitrarily small perturbations, symmetry completely breaks down while entropy remains high. 

It is possible to derive a more stable relationship between information and symmetry by working with words instead. Suppose $\Omega=\{\omega_1,...,\omega_k\}$ and let $P=(p_1,...,p_k)$ be a probability vector.  A word of length $n$ is a function $s:\{1,\dots, n\}\rightarrow \Omega$, and a permutation  $\sigma\in\mathcal S_n$ acts on the set of all these words by precomposition: $s\mapsto s\circ \sigma$.   Provided $np_i$ is an integer for each $i=1,...,k$, the multinomial coefficient 
\begin{equation}
    \binom{n}{nP}=\frac{n!}{(np_1)!\cdots (np_k)!}
\end{equation}
corresponds to the cardinality of the type class $T_P^n$: the words of length $n$ such that $p_i$ represents the relative frequency of the symbol $\omega_i$. The action of $S_n$, restricted to the invariant set $T_P^n$, is transitive, and the stabilizer of any of its elements is isomorphic to $S_{np_1}\times\cdots\times S_{np_k}$, where $np_i$ is the number of occurrences of $\omega_i$. Hence the cardinality of the orbit of a word in this type class is  $\binom n {nP}$, which gives an immediate connection to the Shannon entropy $H(P) =-\sum_{i=1}^k p_i \ln p_i$ via the well-known relation (see e.g. \cite{Varadhan2003})
\begin{equation}\label{eq:limit_binomial}
    \frac 1n \ln \binom n {nP} = H(P) +o(1).
\end{equation}

The second motivating example is given by the action of $\GL_n(\Fq)$ on the total Grassmannian of $\Fq^n$, i.e. the set of all subspaces of $\Fq^n$, as well as the induced action on flags of such spaces (here $q$ is a prime power and $\Fq$ is the field with $q$ elements). Let $V=\Fq^n$ and $\mathcal F_{n,P} = (V_1 \subset\cdots\subset V_k)$ be a flag such that $\dim V_1 = np_1$ and $\dim V_{i} - \dim V_{i-1} = np_i$ for $i=2,...,k$. Such a flag is stabilized by a parabolic subgroup of $\GL_n(\Fq)$, and the cardinality of its orbit $\Orb (\mathcal F_{n,P} ; \GL_n (\Fq))$ is given by the $q$-multinomial coefficient,
\begin{equation}
    \binom{n}{nP} _q =\frac{[n]_q!}{[np_1]_q! \cdots [np_k]_q!},
\end{equation}
where $[k]_q! = (q^k-1)(q^{k-1}-1) \cdots (q^{1}-1)$, provided $k$ is an integer.
In \cite[Prop. 2]{vigneauxfinite}, Vigneaux showed that, asymptotically,
\begin{equation}\label{eq:limit-GL}
   \frac{1}{n^2} \log_q |\Orb (\mathcal F_{n,P} ; \GL_n (\Fq))| =  \Hq (P)+o(1)
\end{equation}
where $\Hq(P) := \frac 12 H_2(P) = \frac 12\big(1-\sum_i p_i^2\big)$ is half the Tsallis $2$-entropy $H_2(P) = 1-\sum_i p_i^2$ (also known as Rao's quadratic entropy, or the Gini--Simpson index). We call $\Hq$ the \emph{(modified) quadratic entropy}, and refer to it simply as the \emph{quadratic entropy} hereafter; using $\Hq$ in place of $H_2$ absorbs a recurring factor of $\frac 12$ in the formulas below.

Our problem is to study the cardinality of quotients of the form observed in these examples, $|G(V)/P (F)|$, where $G$ is a more general finite group and $P$ is a parabolic subgroup. In particular, we consider the cases where there is a finite reflection group or finite group of Lie type, because the symmetric groups $S_n$ form an infinite series of finite reflection groups, and the general linear groups $\GL_n(\Fq)$ a series of  finite groups of Lie type.

\subsection{Main results}

We introduce suitable analogues of the parabolic subgroups that stabilize a word or flag of type given by $P$ as in the examples above, and study quotients of general reflection groups (of the infinite series $B_n$, $C_n$ or $D_n$) and of the corresponding split classical groups over $\Fq$ (e.g. the symplectic group) by these suitable parabolic subgroups.

A uniform treatment of both situations is made possible by classical invariant theory. Every finite reflection group $W$ of rank $n$ has a well-defined multiset of \emph{degrees} $d_1,\dots, d_n$ (Section~\ref{sec:degrees}), which satisfies $|W| = \prod_{i=1}^n d_i$ and, more precisely, \emph{Solomon's formula}: the Poincar\'e polynomial $P_W(t) = \sum_{w\in W} t^{\ell(w)}$ factors as the product of quantum integers $\prod_{i=1}^n [d_i]_t$, where $[d]_t = 1+t+\cdots+t^{d-1}$. For every parabolic subgroup $W_J$ of $W$, the quotient $P_W(t)/P_{W_J}(t)$ is a polynomial that encodes at once the index $|W/W_J|$ (evaluation at $t=1$) and, when $W$ is the Weyl group of a split algebraic group $G$ over $\Fq$ with parabolic subgroup $P_J$, the number of $\Fq$-points of the flag variety $G/P_J$ (evaluation at $t=q$); see \eqref{eq:point_count}. These evaluations are precisely the cardinalities whose asymptotics we study.

By a limiting procedure analogous to \eqref{eq:limit_binomial} and \eqref{eq:limit-GL}, we establish that the entropy associated with reflection groups of type $B_n$, $C_n$ or $D_n$ is $ H_R(P)= H(P)+(1-p_k)\ln 2$, which we call \emph{reflective entropy} (Proposition \ref{prop:reflective_ent}), and the entropy associated with the split groups of Lie type of the same series---the odd orthogonal, symplectic, and split even orthogonal groups---is $\HSp(p_1,...,p_k) = \Hq(P) + \frac 1 2 (1-p_k^2)$, which we call \emph{Lie entropy} (Propositions \ref{prop:lie_entropy} and \ref{prop:spentropy}; the symplectic group is the case we examine in detail). Each one of these entropies has its own ``chain rule'', different from that of Shannon or Tsallis entropies, described by Propositions \ref{prop:chain_rule_reflective} and \ref{prop:chain_rule_sp}; both chain rules are asymptotic shadows of a single identity between Poincar\'e polynomials, \eqref{eq:coarseningdynkin}, evaluated at $t=1$ and at $t=q$ respectively. We also show that the reflective entropy is the Shannon entropy of memoryless sources over \emph{signed} alphabets (Proposition \ref{prop:signed_words}), which extends the classical typicality interpretation of entropy, recalled next, to the hyperoctahedral series.

\subsection{Relation with typical sequences}\label{sec:typicality}

In information theory, entropy is more commonly introduced as the growth rate of sets of \emph{typical sequences}. The link with the perspective adopted here---entropy as growth rate of group orbits---is provided by the method of types of Csisz\'ar and K\"orner \cite{csiszarkorner}; see also the survey \cite{csiszar1998}. The \emph{type} of a word $s\in\Omega^n$ is its empirical distribution $\hat P_s = (n_1/n,\dots,n_k/n)$, where $n_i$ counts the occurrences of $\omega_i$ in $s$. Words of the same type form a single orbit of $S_n$ (a \emph{type class}, the set $T^n_P$ of Section \ref{sec:intro_background}), and the number of types is polynomial in $n$---at most $(n+1)^k$---whereas type classes have exponential size, cf. \eqref{eq:limit_binomial}. Given a memoryless source with marginal law $P$ and a tolerance $\delta>0$, the (strongly) \emph{typical set} $T^n_{[P]\delta}$ consists of the words whose type is $\delta$-close to $P$; the law of large numbers gives $P^{\otimes n}(T^n_{[P]\delta})\to 1$, and by definition $T^n_{[P]\delta}$ is the union of the polynomially many type classes with type near $P$ \cite[Ch.~1.2]{csiszarkorner}. Since polynomial factors do not affect exponential rates,
\begin{equation}\label{eq:typical_set_rate}
    \lim_{\delta\to 0}\ \lim_{n\to\infty}\ \frac 1n \ln |T^n_{[P]\delta}| \,=\, \lim_{n\to\infty} \frac 1n \ln \binom{n}{nP} \,=\, H(P),
\end{equation}
the second limit being taken along the integers $n$ for which $nP$ is integral.
In other words, \emph{up to polynomial factors, the size of the typical set is the size of a single orbit}: probability (concentration) selects the orbit, and the counting of its elements follows from group theory. In Section \ref{sec:signed_typicality} we show that this picture extends verbatim to the reflection groups of types $B$, $C$ and $D$, whose orbits are type classes of words over signed alphabets; Remark \ref{rem:flag_typicality} discusses what survives of it for flag varieties over $\Fq$, and Section \ref{sec:discussion} the analogy with differential entropy and volumes.

\subsection{Notation}

The symbol $q$ denotes a prime power $p^m$ for some $m\in\Z$. The notation $\exp_q(n)=q^n$ is the prime power $q$ raised to the $n$-th power. $\Fq$ denotes the finite field of $q$ elements. A flag $\mathcal F$ in a vector space $V$ is a sequence of subspaces of $V$ which are sequentially contained in each other, and its ``type" is an ordered list of integers $(n_1,\dots, n_k)$ denoting the differences between the dimension of subsequent subspaces. The ``length" of the flag is the number of subspaces it is made of.

Given a finite set $S$, a probability mass function (p.m.f.) or probability distribution is a function $P:S\to [0,1]$ such that $\sum_{s\in S} P(s) = 1$. If the elements of $S$ are totally ordered, such $P$ can be represented as a vector. We move freely between functional and vectorial notations.

\section{Reflection Groups}

\subsection{Basic definitions}

References for this section are \cite{kanereflection} and \cite{humphreysreflection}. Fix an $n$-dimensional Euclidean space $V$. Let $\alpha\in \Rnn$ be a vector and $H_\alpha = \{x\mid (x,\alpha) = 0\}$ be the hyperplane with normal $\alpha$. The associated reflection $s_\alpha$ is the linear map given by the equations:
\begin{align*}
    s_\alpha (\alpha) &= -\alpha, \\
    s_\alpha (x) &= x\quad \mbox{if } x\in H_\alpha
\end{align*}

A group $W$ is a reflection group if it is generated by a set of reflections. A root system $\Phi\subset V$ for a finite reflection group $W$ is a finite set of non-zero vectors such that \cite[p. 26]{kanereflection}
\begin{enumerate}
    \item For any $\alpha \in \Phi$, $\lambda \alpha \in\Phi$ if and only if $\lambda = \pm 1$. 
  \item For any $\alpha, \beta\in \Phi$, $s_\alpha( \beta) \in \Phi$.   
\end{enumerate}
 Each element of $\Phi$ is called a root. A fundamental root system is a subset $\Delta\subset \Phi$ such that $\Delta$ is linearly independent and every element of $\Phi$ is a linear combination of elements of $\Delta$ where the coefficients are either all nonnegative or all nonpositive.

 The Poincaré polynomial of a subgroup $W_X < W$ of a group can be defined as \[P_{W_X}(t) = \sum_{g\in W_X} t^{\ell (g)}\]
where $\ell(g)$ denotes the number of positive roots that are sent to negative roots by $g$ in $W_X$ (or $W$); see \cite[Secs. 1.6-1.11]{humphreysreflection}. When the Poincaré polynomial is evaluated at 1, it gives the cardinality of the subgroup; $P_{W_X}(1)=|W_X|$.

Finite reflection groups are completely classified by their fundamental root systems, and more importantly, the Dynkin diagram associated to any of them. Each node of the Dynkin diagram represents a fundamental root of a chosen fundamental root system $\Delta$. Two nodes are connected by an edge if their actions do not commute or, equivalently, if the smallest integer $m$ such that $(s_\alpha s_\beta)^m = 1$ is greater than 2. The $B_n$ and $C_n$ root systems lead to the same cardinality of reflection group, $2^n n!$, whereas the $D_n$ root system leads to  $2^{n-1} n!$.

\begin{table}[htbp]
    \renewcommand{\arraystretch}{1.3}
    \caption{Corresponding Root Systems, Groups, Dynkin Diagrams, and Degrees (see Section \ref{sec:degrees})}
    \label{tab:dynkin}
    \centering
    \begin{tabular}{c|c|c|c}
            \hline
                R.S. & Group & Dynkin Dgrm. & Degrees  \\
                \hline
                \hline
                $A_n$ & $S_{n+1}$ & \dynkin[scale=1.5] A{} & $2,3,\dots,n{+}1$ \\
                \hline
                $B_n$ & $\Z_2^n \rtimes S_n$ & \dynkin[scale=1.5] B{} & $2,4,\dots,2n$ \\
                \hline
                $C_n$ & $\Z_2^n \rtimes S_n$ & \dynkin[scale=1.5] C{} & $2,4,\dots,2n$ \\
                \hline
                $D_n$ & $\Z_2^{n-1} \rtimes S_n$ & \dynkin[scale=1.5] D{} & $2,4,\dots,2n{-}2;\, n$ \\
                \hline
    \end{tabular}
\end{table}

Parabolic subgroups are the subgroups generated by the Dynkin diagrams that result from excluding a subset of the fundamental roots. Because reflections in two different connected components of the diagram commute, parabolic subgroups are products of reflection groups of type $A_n$, $B_n$, $C_n$, and $D_n$, and their Poincar\'e polynomials are computed as follows.

\begin{proposition}\label{prop:factorization_poincare_pol}
    If $W_X$ and $W_Y$ are parabolic subgroups of $W$ which intersect trivially and for which any $x\in W_X$ and $y\in W_Y$ commute ($(xy)^2=1$), then the Poincar\'e polynomial of the product of these groups is the product of the Poincar\'e polynomials of them individually:
\begin{equation}\label{eq:factoringpoincare}
    P_{W_X}(t) P_{W_Y}(t) = P_{W_X W_Y}(t).
\end{equation}
\end{proposition}
\begin{proof}
    See Appendix \ref{app:factorpoincare}.
\end{proof}

\subsection{Invariant theory, degrees, and Solomon's formula}\label{sec:degrees}

There is a general product formula for the Poincar\'e polynomial of any finite reflection group and, more generally, for the quotient of Poincar\'e polynomials attached to any parabolic subgroup. It comes from invariant theory. Let $W$ be a finite reflection group acting on $V\cong \Rnn$, and let $R=\mathrm{Sym}(V^*)$ be the algebra of polynomial functions on $V$, on which $W$ acts by precomposition. A theorem of Chevalley \cite{chevalley1955} (see also \cite[Ch.~3]{humphreysreflection}) states that the subalgebra $R^W$ of $W$-invariant polynomials is itself a polynomial algebra, freely generated by $n$ algebraically independent homogeneous polynomials $f_1,\dots,f_n$. The generators are not unique, but their degrees $d_i = \deg f_i$ are uniquely determined; they are called the \emph{degrees} of $W$. The degrees are explicitly known for each irreducible type \cite[Sec.~3.7]{humphreysreflection}; those of the infinite series appear in the last column of Table \ref{tab:dynkin}.

Two classical facts relate the degrees to the enumeration problems studied here. First \cite[Sec.~3.9]{humphreysreflection},
\begin{equation}\label{eq:prod_degrees}
    |W| = d_1 d_2 \cdots d_n,
\end{equation}
and the number $N$ of positive roots of $W$ equals $\sum_{i=1}^n (d_i-1)$. Second, \emph{Solomon's formula} \cite{solomon1966}, \cite[Sec.~3.15]{humphreysreflection}: the Poincar\'e polynomial of $W$ is the $q$-deformation of \eqref{eq:prod_degrees} given by the product of \emph{quantum integers} $[d]_t := 1+t+\cdots+t^{d-1}$,
\begin{equation}\label{eq:solomon}
    P_W(t) = \prod_{i=1}^n\, [d_i]_t.
\end{equation}
Since $[d]_1 = d$, the evaluation of \eqref{eq:solomon} at $t=1$ recovers \eqref{eq:prod_degrees}; the degree of $P_W$ is $N=\sum_i (d_i - 1)$.

For $J\subset \Delta$, the parabolic subgroup $W_J$ is itself a reflection group: the product of the irreducible reflection groups corresponding to the connected components of the subdiagram spanned by $J$ (cf. Proposition \ref{prop:factorization_poincare_pol}). Hence $P_{W_J}$ is also given by \eqref{eq:solomon}, in terms of the degrees of $W_J$. Moreover, every coset in $W/W_J$ has a unique representative of minimal length, and denoting by $W^J$ the set of such representatives one has \cite[Sec.~1.10]{humphreysreflection}, \cite[Sec.~2.4 \& Ch.~7]{bjornerbrenti}:
\begin{equation}\label{eq:poincare_quotient}
    \frac{P_W(t)}{P_{W_J}(t)} = P_{W^J} = \sum_{w\in W^J} t^{\ell(w)},
\end{equation}
a \emph{polynomial} with nonnegative integer coefficients, of degree $N(W)-N(W_J)$. At $t=1$ it counts the cosets of $W_J$ in $W$; its values at prime powers $t=q$ acquire a geometric meaning in Section \ref{sec:lie}.

\subsection{Asymptotic Growth of Quotients}\label{sec:symptotic_reflection_groups}

Let $P=(p_1,...,p_k)$ be a probability vector with rational entries. We consider in this section a set $N_P$ of positive integers $n$ such that  $np_i\in\mathbb Z$ and $np_i> 3$ for  $i=1,...,k$. 

Given a reflection group $W$ of type $A_{n-1}$, $B_{n-1}$, $C_{n-1}$ or $D_{n-1}$ with Dynkin diagram as in Table \ref{tab:dynkin}, with $n\in N_P$, we define the \emph{parabolic subgroup $W_P$ associated with the probability vector P}  as the parabolic subgroup corresponding to excluding the $np_1$-th root, the $n(p_1+p_2)$-th root, ... through the $n(p_1+\cdots+ p_{k-1})$-th root from the diagram. (We do not want to remove the last root in the diagram, because this would reduce all cases to the $A_n$ case.)

If $W$ is of type $A_{n-1}$, then $W\cong S_n$ and $W_P$ is isomorphic to $S_{np_1}\times \cdots \times S_{np_k}$. As we explained in the introduction, 
\begin{equation}\label{eq:entropy-from-parabolic}
 \left|W/W_P\right| =\frac{|W|}{|W_P|}= {n\choose nP} = \exp(nH(P)+o(n)).
\end{equation}

\begin{proposition}\label{prop:reflective_ent}
    Let $P$ be a probability vector,  $W$  a finite reflection group of type $B_{n-1}$, $C_{n-1}$ or $D_{n-1}$ with $n\in N_P$, and $W_P$ the parabolic subgroup associated with $P$. In this setting,
\begin{equation}\label{eq:new-entropy-reflections}
\lim_{\substack{n\rightarrow\infty \\ n\in N_P }} \frac 1 n \ln |W/W_P| =  H(P)+(1-p_k)\ln 2.
\end{equation}
\end{proposition}
\begin{proof}
    We prove the claim for $B_{n-1}$, the other cases being analogous. We then have $W\cong \Z_2^{n-1}\rtimes S_{n-1}$ and $W_P \cong S_{np_1}\times \cdots\times S_{np_{(k-1)}}\times \Z_2^{np_k-1}\rtimes S_{np_k-1}$, as one can see from the diagram 
\begin{equation*}
        W_P \leftrightarrow \dynkin[labels={,,np_1,,,np_{k-1}}, scale=1.5] B{*.*t*..*t*.**}. 
\end{equation*} 
Therefore, computing the orders via the product of the degrees \eqref{eq:prod_degrees} and Table \ref{tab:dynkin},
\begin{equation}
    \frac{|W|}{|W_P|} = \frac{2^{n-1}\times (n-1)!}{np_1! \times\cdots\times np_{k-1}! 2^{np_k-1}\times (np_k-1)!}
    \end{equation}
    from which the result easily follows.
\end{proof}

We call 
\begin{equation}
    H_R(p_1,...,p_k) = H(P)+(1-p_k)\ln 2
\end{equation}
the \emph{reflective entropy} of the probability vector $P$.

\subsection{Typicality: words over signed alphabets}\label{sec:signed_typicality}

The typicality picture of Section \ref{sec:typicality} extends to the series $B$, $C$ and $D$: the quotients $W/W_P$ are again sets of words, over a partially \emph{signed} alphabet, and the reflective entropy is again a Shannon entropy.

Let $\Omega = \{\omega_1^+,\omega_1^-,\dots,\omega_{k-1}^+,\omega_{k-1}^-,\omega_k\}$ be an alphabet in which the first $k-1$ symbols carry a sign and the last one is neutral; for $i<k$ we write $|\omega_i|$ for the common magnitude of $\omega_i^+$ and $\omega_i^-$. The hyperoctahedral group $\Z_2^n\rtimes S_n$ (the reflection group of type $B_n$, or equivalently $C_n$) acts on the set of words $\Omega^n$: the symmetric group permutes the positions, and the $i$-th factor $\Z_2$ switches the sign of the letter at position $i$ (with the convention $\omega_k^\pm := \omega_k$). Call \emph{unsigned type} of a word the vector $(m_1,\dots,m_k)$ counting the occurrences of the magnitude $|\omega_i|$ (that is, of $\omega_i^+$ or $\omega_i^-$), for $i<k$, and of $\omega_k$.

\begin{proposition}[Signed words]\label{prop:signed_words}
    Let $W = \Z_2^n \rtimes S_n$ act on $\Omega^n$ as above, and let $P=(p_1,\dots,p_k)$ be a probability vector such that $nP$ is integral. Then:
    \begin{enumerate}
        \item The words of unsigned type $nP$ form a single $W$-orbit, of cardinality $\binom{n}{nP}\, 2^{n(1-p_k)}$.
        \item The stabilizer of any such word is a parabolic subgroup of $W$ isomorphic to $S_{np_1}\times\cdots\times S_{np_{k-1}}\times (\Z_2^{np_k}\rtimes S_{np_k})$, i.e. of the shape considered in Proposition \ref{prop:reflective_ent}; the orbit is thus identified with a quotient $W/W_P$.
        \item Let $\tilde P$ be the sign-symmetric law on $\Omega$ that assigns mass $p_i/2$ to each of $\omega_i^\pm$, for $i<k$, and $p_k$ to $\omega_k$. Then
        \begin{equation}\label{eq:HR_is_Shannon}
            H(\tilde P) = H(P) + (1-p_k)\ln 2 = H_R(P).
        \end{equation}
    \end{enumerate}
\end{proposition}
\begin{proof}
    (1) Transitivity is clear: sign switches act independently at each position and permutations rearrange magnitudes; the words of unsigned type $nP$ are counted by choosing the positions of each magnitude ($\binom{n}{nP}$ options) and then a sign for each of the $n(1-p_k)$ signed letters. (2) Consider the sorted word $s^* = (\omega_1^+,\dots,\omega_1^+,\omega_2^+,\dots,\omega_k)$. An element $(\epsilon,\sigma)\in W$ fixes $s^*$ if and only if $\sigma$ preserves the blocks of constant magnitude and $\epsilon$ is trivial at the signed positions; at the $np_k$ neutral positions $\epsilon$ is arbitrary. This subgroup is generated by the simple reflections of $W$ other than those at positions $np_1, n(p_1{+}p_2),\dots,n(1{-}p_k)$ of the Dynkin diagram, the sign switches being attached to the last block. (3) Direct computation:
    $H(\tilde P) = -\sum_{i<k} 2\,\frac{p_i}2 \ln \frac{p_i}2 - p_k \ln p_k = H(P) + (1-p_k)\ln 2$.
\end{proof}

Consequently, the whole method of types applies. Let $X_1,\dots,X_n$ be drawn from a memoryless source on $\Omega$ whose marginal law is \emph{sign-symmetric}, i.e. invariant under $\omega_i^+\leftrightarrow\omega_i^-$; equivalently, the law $\tilde P$ of Proposition \ref{prop:signed_words} for some probability vector $P$. By the law of large numbers, the unsigned empirical type concentrates around $P$; the set of words with unsigned type $\delta$-close to $P$ is a union of polynomially many $W$-orbits; and each $W$-orbit is itself a union of at most $(n+1)^{k-1}$ ordinary type classes over $\Omega$ (one for each splitting of the $m_i$ letters of magnitude $|\omega_i|$ into $+$ and $-$), among which the balanced splitting is the largest. All these sets therefore share the exponential growth rate $H(\tilde P) = H_R(P)$: \emph{up to polynomial factors, the typical set of a sign-symmetric source is a single orbit of the hyperoctahedral group}, and the reflective entropy is the Shannon entropy of such sources. For the group of type $D_n$ (even numbers of sign switches), each orbit splits into at most two suborbits, according to the parity of the number of minus signs, which leaves the exponential rates unchanged---in agreement with Proposition \ref{prop:reflective_ent}.

\subsection{Understanding Coarsening with Dynkin Diagrams}\label{sec:coarsening}

In information-theoretic applications, a probability vector $P=(p_1, \dots, p_k)$ can be regarded as the law of a discrete random variable $X$ with possible outcomes $E_X=\{1,\dots, k\}$. Coarse-graining refers to mapping this outcome space onto a smaller set $E_Y$ by $\pi:E_X\to E_Y$. The corresponding push-forward $\pi_* P$ of the law $P$, given by
\begin{equation}
    \pi_* P(y) = \sum_{i\in \pi\inv (y)} p_i
\end{equation}
can be regarded as the law of an induced random variable $Y$. For each $y\in E_Y$, there is a conditional probability distribution $P \mid _{Y=y}$ given by
\begin{equation}
    P\mid_{Y=y}(x) = \begin{cases}
        P(x)/\pi_*P(y) & \text{if } x\in \pi^{-1}(y)\\
        0 & \text{otherwise}
    \end{cases}.
\end{equation}
One can verify directly that Shannon entropy satisfies the chain rule 
\begin{equation}\label{eq:chain_rule_ent}
    H(P) = H(\pi_*P) + \sum_{y\in E_Y} \pi_*P(y) H(P|_{Y=y}).
\end{equation}
Interestingly, this identity also follows from an asymptotic expansion of the multinomial coefficients in the elementary identity
\begin{equation}\label{eq:mult_chain_rule}
    { n\choose nP} = {n\choose n(\pi_*P) } \prod_{y\in E_Y }{n\pi_*P(y)  \choose n P|_{Y=y} }.
\end{equation}
Hence the entropic chain rule \eqref{eq:chain_rule_ent} is an ``asymptotic shadow'' of the multinomial chain rule \eqref{eq:mult_chain_rule}. In \cite{vigneauxfinite}, Vigneaux found an interpretation along these lines for the deformed chain rule of Tsallis 2-entropy; this entropy describes the asymptotic growth of $q$-multinomial coefficients as we mentioned in the introduction. 

In this section, we extend this structural connection to reflection groups. We proceed in two steps:
\begin{itemize}
    \item First, by noting that there is a generalization of \eqref{eq:mult_chain_rule} that follows from the general properties of the Poincaré polynomial presented in Section \ref{sec:degrees}.
    \item Second, by combining this identity with the asymptotic expansion \eqref{eq:new-entropy-reflections} one might obtain a purely combinatorial account of the chain rule that $H_R$ obeys. 
\end{itemize}
We refer to these as the combinatorial and probabilistic steps, respectively. 

\subsubsection{Combinatorial step}

Let $\Delta$ be a fundamental root system with $n$ roots (ordered), corresponding reflection group $W(\Delta)$ and Dynkin diagram $\mathbf D$.  Let $I'$ and $J'$ be subsets of nonconsecutive roots and let $I$ and $J$ be their respective complements in $\Delta$,  such that $I\subset J\subset \Delta$. We say that $i,j\in \Delta$ are connected if there is an edge $(i,j)$ in $\mathbf D$;  let $\mathbb I$ be the set of subsets of $I$ which are connected components in $\mathbf D$, and that are not subsets of any other  connected component, and likewise let $\mathbb J$ be the set of subsets of $J$ which are connected components in $\mathbf D$ and that are not subsets of any other  connected component. (That is: $\mathbb I$ are the pieces that remain of $\mathbf D$ once the roots $I'$ are removed, etc.)  

\begin{proposition}
    Given the objects described above,
    \begin{equation}\label{eq:coarseningdynkin}
        \frac{P_W(t)}{P_{W_I}(t)} =\frac{P_W(t)}{P_{W_J}(t)}\cdot \prod_{\substack{S\in \mathbb J\\ S\not\in \mathbb I}}\frac{P_{W_S}(t)}{P_{W_{S\cap I}}(t)}
    \end{equation}
\end{proposition}

\begin{proof}
    We use Proposition \ref{prop:factorization_poincare_pol} and the fact that the elements of $\mathbb J$ which are shared with $\mathbb I$ could be factored out of both $P_{W_J}(t)$ and $P_{W_I}(t)$ since they commute with the parts of $\mathbb J$ and $\mathbb I$ which are not shared, leaving us with:  $$\frac{P_{W_J}(t)}{P_{W_I}(t)}=\prod_{\substack{S\in \mathbb J\\ S\not\in \mathbb I}}\frac{P_{W_S}(t)}{P_{W_{S\cap I}}(t)}.$$
\end{proof}

When the Poincaré polynomial is evaluated at $1$, we recover an identity involving cardinalities:
\begin{equation}\label{eq:cardinalities_coarse_grained_reflection}
    \frac{|W|}{|W_I|} = \frac{|W|}{|W_J|} \cdot \prod_{\substack{S\in \mathbb J\\ S\not\in \mathbb I}}\frac{|W_S|}{|W_{S\cap I}|},
\end{equation}
which in the particular case of reflective groups of type $A_n$ reduces to an identity of multinomial coefficients of which \eqref{eq:mult_chain_rule} is a particular case.

We emphasize that \eqref{eq:coarseningdynkin} is an identity between \emph{polynomials} in $t$. Besides the evaluation at $t=1$ just described, one can evaluate it at a prime power $t=q$; in view of \eqref{eq:point_count} below, it then becomes an identity between numbers of points of flag varieties over $\Fq$, which we use in Section \ref{sec:asymptotics_symplectic} to derive the chain rule of the Lie entropy.

\subsubsection{Probabilistic step}

In particular, we set $E_Y = \{1,\dots, m\}$ and impose that $\pi$ is an increasing surjection. We introduce again the set $N_P$ of positive integers   $n$ such that $np_i\in \Z$ for $i=1,\dots, k$. Denote by $I_P'$ the set $\{np_1,\dots, np_k\}$ and by $J_P'$ the set $\{n\pi_* P(1),\dots, n\pi_* P(m)\}$.

Let $\Delta$ be a fundamental root system with $n\in N_P$ roots, corresponding reflection group $W(\Delta)$ and Dynkin diagram $\mathbf D$. We consider, for $I_P= (I_P')^c \subset \Delta$, the parabolic subgroups $W_{I_P}$ corresponding to removing fundamental roots only at locations of elements of $I_P'$ and $W_{J_P}$ corresponding to removing fundamental roots only at locations of elements of $J_P'$. 

For $j\in E_Y$, we introduce the conditional probabilities on $E_X$:
$$P|_{\pi^{-1}(j)} = \begin{cases}
    \frac{p_i}{\pi_*P(j)} & \text{for } i\in \pi^{-1}(j)\\
    0 & \text{otherwise}
\end{cases}  $$
Remark that connected components of roots in $J_P$ in the Dynkin diagrams that are also connected components of $I_P$ are associated with a conditional probability that is a Dirac measure and hence has vanishing Shannon entropy. 

\begin{proposition}\label{prop:chain_rule_reflective}
    Let $P$ be a probability vector as above and $Q$ its push-forward under $\pi:E_X\to E_Y$. Then, 
    \begin{equation*}
        H_R(P) = H_R(Q) + \sum_{j=1}^{m-1} Q(j) H(P|_{\pi^{-1}(j)}) + Q(m) H_R(P|_{\pi^{-1}(m)}). 
    \end{equation*}
\end{proposition}
\begin{proof}
    We apply logarithms to \eqref{eq:cardinalities_coarse_grained_reflection}, multiply by $\frac 1n$, and then take the limit $n\to \infty$ with $n\in N_P$ According to Proposition \ref{prop:reflective_ent}, the limit is  $H_R(P)$ for $|W|/|W_I|$, $H_R(Q)$ for $|W|/|W_J|$, and $Q(m)H_R(P|_{\pi^{-1}(m)})$ for $|W_{S^*}|/|W_{S^*\cap I}|$ where $S^*$ is the only connected set of roots in the Dynkin diagram that contains the rightmost, non-$A_n$-like part. The other quotients $|W_S|/|W_{S\cap I}|$ are $A_n$-like and their limit is computed as in \eqref{eq:limit_binomial}.
\end{proof}

\section{Finite Groups of Lie Type}\label{sec:lie}

\subsection{Flag varieties over $\Fq$ and Poincaré polynomials}\label{sec:flag_varieties}

The reflection groups of the series $A$--$D$ arise as Weyl groups of the split classical algebraic groups over $\Fq$: the general linear group $\GL_n(\Fq)$ has Weyl group of type $A_{n-1}$, the odd orthogonal group $\SO_{2n+1}(\Fq)$ of type $B_n$, the symplectic group $\Sp_{2n}(\Fq)$ of type $C_n$, and the split even orthogonal group $\SO^+_{2n}(\Fq)$ of type $D_n$ \cite{carter}. In each case, the subsets $J$ of the fundamental root system $\Delta$ classify the \emph{standard parabolic subgroups} $P_J$ of $G$; for instance, $P_\Delta = G$, and $P_\emptyset = B$ is a \emph{Borel subgroup} (e.g. the upper triangular matrices in $\GL_n$). The quotient $G/P_J$ is a \emph{flag variety}: for the classical groups, its points are flags of subspaces of the natural module---arbitrary subspaces in type $A$, totally isotropic subspaces with respect to the defining bilinear form in types $B$, $C$ and $D$---whose dimensions are prescribed by the removed nodes $\Delta\setminus J$.

The Bruhat decomposition $G = \bigsqcup_{w\in W} BwB$ induces a decomposition of $G/P_J$ into affine cells (Schubert cells) indexed by the minimal-length representatives $w\in W^J$, the cell of $w$ being an affine space of dimension $\ell(w)$; see \cite{carter,garrettbuildings}. Counting $\Fq$-points cell by cell and invoking \eqref{eq:poincare_quotient} and Solomon's formula \eqref{eq:solomon}, one obtains the general product formula \cite[\S2.9, p.~75]{carter}
\begin{equation}\label{eq:point_count}
    |(G/P_J)(\Fq)| \,=\, \sum_{w\in W^J} q^{\ell(w)} \,=\, \frac{P_W(q)}{P_{W_J}(q)} \,=\, \frac{\prod_{i=1}^n [d_i]_q}{P_{W_J}(q)}.
\end{equation}
Thus the number of $\Fq$-points of the flag variety $G/P_J$ is a $q$-deformation of the index $|W/W_J| = P_W(1)/P_{W_J}(1)$ studied in the previous section: the two enumeration problems are governed by the same polynomial, evaluated at $t=q$ and $t=1$ respectively.

\emph{Example (type $A$).} The degrees of $A_{n-1}$ are $2,3,\dots,n$, so $P_{S_n}(q) = \prod_{i=1}^{n} [i]_q$. If $J$ is obtained from $\Delta$ by removing the nodes at positions $np_1, n(p_1+p_2),\dots, n(1-p_k)$, then $W_J \cong S_{np_1}\times\cdots\times S_{np_k}$ and \eqref{eq:point_count} gives
\begin{equation*}
    |(\GL_n/P_J)(\Fq)| = \frac{\prod_{i=1}^{n}[i]_q}{\prod_{j=1}^{k}\prod_{i=1}^{np_j}[i]_q} = \binom{n}{nP}_q,
\end{equation*}
the number of flags of type $nP$ in $\Fq^n$ that appears in \eqref{eq:limit-GL}. (Since $q^i - 1 = (q-1)[i]_q$ and $\sum_j np_j = n$, the factors $(q-1)$ cancel and both conventions for the $q$-factorials give the same quotient.)

\begin{remark}[Topological meaning]\label{rem:betti}
    The same polynomial \eqref{eq:point_count} describes the topology of the complex points: $G(\mathbb C)/P_J(\mathbb C)$ is a compact complex manifold whose odd Betti numbers vanish and whose even Betti numbers are $b_{2i} = |\{w\in W^J \mid \ell(w) = i\}|$; see \cite{bgg1973}. This follows from the Schubert cell decomposition already invoked: the cells are complex affine spaces, hence even-dimensional as real manifolds, so the cellular chain complex has trivial differentials and the (co)homology is free with the stated ranks. In particular, the Euler characteristic of $G(\mathbb C)/P_J(\mathbb C)$ equals the total Betti number $\sum_i b_{2i} = |W/W_J|$. The entropies studied in this article thus admit a topological reading: the reflective entropy measures the exponential growth rate of Euler characteristics of flag manifolds, while the Lie entropy introduced below measures the growth rate of their numbers of points over $\Fq$.
\end{remark}

\begin{remark}[Types $B$ and $C$]\label{rem:BC}
    The Coxeter groups of types $B_n$ and $C_n$ coincide, with the same parabolic subgroups, the same degrees, and the same Poincaré polynomials; the two root systems differ only in the lengths of their roots. The associated root data---and hence the algebraic groups $\SO_{2n+1}$ and $\Sp_{2n}$---are different, and the corresponding flag varieties are \emph{not} isomorphic. Nevertheless, by \eqref{eq:point_count}, they have exactly the same number of points over every finite field. From the counting perspective adopted in this article, types $B$ and $C$ are therefore indistinguishable.
\end{remark}

\subsection{Entropy asymptotics for the series $B$, $C$ and $D$}\label{sec:lie_asymptotics}

The general formula \eqref{eq:point_count} yields at once the asymptotic behavior of the number of points of the flag varieties associated with the three series $B_n$, $C_n$, $D_n$, generalizing \eqref{eq:limit-GL}.

\begin{proposition}\label{prop:lie_entropy}
    Let $P=(p_1,\dots,p_k)$ be a probability vector with rational entries and $N_P$ as in Section \ref{sec:symptotic_reflection_groups}. For $n\in N_P$, let $G_n$ be a split classical group over $\Fq$ whose Weyl group $W$ is of type $B_n$, $C_n$ or $D_n$---that is, $G_n = \SO_{2n+1}$, $\Sp_{2n}$ or $\SO^+_{2n}$---and let $P_J$ be the standard parabolic subgroup obtained by removing the nodes at positions $np_1, n(p_1+p_2), \dots, n(1-p_k)$ of the Dynkin diagram. Write $X\in\{B,C,D\}$ for the type of $W$, and set $e_B=e_C=1,e_D=0$. Then
    \begin{equation} \label{eq:quad_lin}
        \begin{split}
            \log_q |(G_n/P_J)(\Fq)|
            = n^2\left(\Hq(P)+\frac12(1-p_k^2)\right)
            +\left(e_X-\frac12\right)n(1-p_k)+O_{q,k}(1).
        \end{split}
    \end{equation}
    In particular,
    \begin{equation*}
        \lim_{\substack{n\to\infty\\ n\in N_P}} \frac{1}{n^2}\log_q |(G_n/P_J)(\Fq)| = \Hq(P) + \frac 12 (1-p_k^2).
    \end{equation*}
\end{proposition}
\begin{proof}
    The polynomial $P_W(t)/P_{W_J}(t) = \sum_{w\in W^J} t^{\ell(w)}$ has nonnegative integer coefficients and degree $D_n := N(W)-N(W_J)$, where $N(\cdot)$ denotes the number of positive roots. By Solomon's formula and the identity
    \[
        [d]_q = q^{d-1}\frac{1-q^{-d}}{1-q^{-1}},
    \]
    the explicit degree lists in Table \ref{tab:dynkin} give 
    \begin{equation} \label{eq:point_count_degree_error}
        \log_q |(G_n/P_J)(\Fq)| = D_n+O_{q,k}(1).
    \end{equation}
    Since $W$ has rank $n$, whereas $W_J$ has rank $n-k+1$, the factors $(1-q^{-1})^{-1}$ leave only a fixed power depending on $k$. The remaining factors are uniformly bounded because
    \[
        \sum_{d\ge2} \left| \log_q(1-q^{-d}) \right| < \infty
    \] and $W_J$ has only $k$ connected components.
    By Section \ref{sec:degrees} and Table \ref{tab:dynkin}, $N(W) = \sum_i (d_i-1)$ equals $n^2$ in types $B_n$ and $C_n$, and $n(n-1)$ in type $D_n$. The parabolic subgroup $W_J$ is of type $A_{np_1-1}\times\cdots\times A_{np_{k-1}-1}\times X_{np_k}$, where $X=B$, $C$ or $D$ respectively, so that $N(W_J) = \sum_{i=1}^{k-1}\binom{np_i}{2} + N(X_{np_k})$, with $N(X_{np_k}) = (np_k)^2$ or $np_k(np_k-1)$. Equivalently,
    \[
        N(X_r) = r^2+(e_X-1)r.
    \] Therefore, using $\sum_{i=1}^{k-1}p_i=1-p_k$,
    \begin{align*}
        D_n &= N(X_n) -\sum_{i=1}^{k-1}\binom{np_i}{2} -N(X_{np_k}) \\
            &= n^2\left( 1-\frac12\sum_{i=1}^{k-1}p_i^2-p_k^2 \right) + \left(e_X-\frac12\right)n(1-p_k) \\
            &= n^2\left( \Hq(P)+\frac12(1-p_k^2) \right) + \left(e_X-\frac12\right)n(1-p_k).
    \end{align*}
    Substituting in $\eqref{eq:point_count_degree_error}$, it yields the desired claim.
\end{proof}

The same argument in type $A_{n-1}$, where $N(W) = \binom n2$ and the parabolic has no distinguished tail, gives $D_n=n^2\Hq(P)$ exactly. Thus there is no linear term and we recover the limit $\Hq(P)$ of \eqref{eq:limit-GL}. Note also that all three series $B,C,D$ lead to a same leading entropic functional, exactly as the reflection groups of types $B$, $C$ and $D$ all lead to the reflective entropy; for types $B$ and $C$ this is clear from Remark \ref{rem:BC}, since even the counts at finite $n$ agree.

We call
\begin{equation}
    \HSp(p_1,...,p_k) = \Hq(P) + \frac 1 2 (1-p_k^2)
\end{equation}
the \emph{Lie entropy} of the probability distribution $P = (p_1,...,p_k)$. 

The same sign-symmetric law $\tilde P$ of Proposition~\ref{prop:signed_words} also realizes Lie entropy. Indeed,
\begin{equation}
    \HSp(P) = 1 - \frac12 \sum_{i=1}^{k-1}p_i^2-p_k^2
    = H_2(\tilde P) = 2\Hq(\tilde P)
\end{equation}

\begin{remark}[Polar spaces and twisted forms]\label{rem:polar}
    The coincidence goes further. The lattices of totally isotropic subspaces of the classical forms are axiomatized by the \emph{finite classical polar spaces}: besides the symplectic space $W(2n-1,q)$ (type $C_n$) and the quadrics $Q(2n,q)$ (type $B_n$) and $Q^+(2n-1,q)$ (type $D_n$), the list comprises the elliptic quadric $Q^-(2n+1,q)$ and two Hermitian spaces $H(2n-1,q^2)$ and $H(2n,q^2)$, whose isometry groups are \emph{twisted} groups of Lie type ($^2D_{n+1}$ and $^2A$), not covered by the product formula \eqref{eq:point_count} for split groups. There is nonetheless a uniform count \cite[Lemma~9.4.1]{bcn}, \cite[Lem.~2.10]{deboecketal}: a polar space of rank $n$ with parameter $e$ contains
    \begin{equation*}
        \binom{n}{s}_q\, \prod_{i=1}^{s} \big(q^{\,n+e-i}+1\big)
    \end{equation*}
    totally isotropic subspaces of dimension $s$, where $q$ is the order of the base field (a square, for the Hermitian spaces) and $e = 0, \frac 12, 1, 1, \frac 32, 2$ for $Q^+$, $H(2n-1,q^2)$, $W$, $Q$, $H(2n,q^2)$, $Q^-$ respectively; see also \cite[Table~1]{deboecketal}. Writing $s = \sigma n$, the logarithm in base $q$ of this count equals
    \begin{equation*}
        \Big(2\sigma - \frac 32 \sigma^2\Big)\, n^2 + \Big(e - \frac 12\Big)\,\sigma\, n + O(1).
    \end{equation*}
    More generally, set $m_i=np_i$ and $s=n-m_k$. An isotropic flag of type $(m_1,\dots,m_{k-1})$ can be obtained by first choosing its terminal isotropic $s$-subspace and then choosing an ordinary flag inside it. Hence we can count it by
    \[
        \binom{n}{s}_q \prod_{i=1}^{s}(q^{\,n+e-i}+1) \binom{s}{m_1,\dots,m_{k-1}}_q = \binom{n}{m_1,\dots,m_k}_q \prod_{i=1}^{n-m_k}(q^{\,n+e-i}+1).
    \]
    Consequently,
    \[
         \log_q\#\{\text{isotropic flags of type }nP\} = n^2\HSp(P) +\left(e-\frac12\right)n(1-p_k) +O_{q,k}(1).
    \]
    Thus the two-term formula~\eqref{eq:quad_lin} extends to arbitrary flags in all six polar-space families. The preceding one-subspace formula is the binary case $k=2$. The leading term $\HSp(P)$ is common to all six families, while the linear term records the parameter $e$. In particular, types $B$ and $C$ share even this term (both have $e=1$), sharpening Remark \ref{rem:BC}, whereas $D$ ($e=0$), $^2D$ ($e=2$) and the Hermitian forms are distinguished by their order-$n$ term.
\end{remark}

\subsection{Example: the symplectic group and isotropic flags}\label{sec:symplectic}

We now describe concretely the flag varieties of type $C_n$. A \emph{symplectic form} on a vector space $V$ over $\Fq$ is a nondegenerate bilinear function $f:V\times V\to\Fq$ such that $f(y,x)=-f(x,y)$ and $f(x,x)=0$ for all $x,y\in V$; the pair $(V,f)$ is then called a \emph{symplectic vector space}, and its dimension is necessarily even, say $2n$. All symplectic forms on $\Fq^{2n}$ are equivalent to a standard one \cite[Ch.~2]{classicalgrove}, and the \emph{symplectic group} $\Sp(V)\cong \Sp_{2n}(\Fq)$ consists of the maps $g\in\GL(V)$ such that $f(gx,gy)=f(x,y)$ for all $x,y\in V$. A subspace $U\subset V$ is (totally) \emph{isotropic} if $f|_{U\times U}\equiv 0$, in which case $\dim U\leq n$. An \emph{isotropic flag} of type $(m_1,\dots,m_{k-1})$ is a chain $\flis F = (S_1\subset\cdots\subset S_{k-1})$ of isotropic subspaces of $V$ with $\dim S_i = \sum_{j=1}^i m_j$.\footnote{ In the analogy with words, as in Section \ref{sec:intro_background}, the case $k=2$ corresponds to binary sequences and in general $k$ can be seen as the number of symbols used.}

The group $\Sp(V)$ acts transitively on the isotropic flags of a given type, and the stabilizer $P(\flis F) = \{g\in\Sp(V)\mid gS_i = S_i, \ i=1,\dots,k-1\}$ of a flag of type $(np_1,\dots,np_{k-1})$ is (a conjugate of) the standard parabolic subgroup $P_J$ of Proposition \ref{prop:lie_entropy}; see Appendix \ref{app:symplectic} for elementary proofs. The quotient counted by \eqref{eq:point_count} is therefore in bijection with the set of isotropic flags of type $nP$.

In type $C_n$, moreover, the general product formula becomes fully explicit. The degrees are $2,4,\dots,2n$ and the quantum integers factor as $[2i]_q = [i]_q\,(q^i+1)$, so Solomon's formula \eqref{eq:solomon} reads
\begin{equation*}
    P_W(q) = \prod_{i=1}^n [2i]_q = \prod_{i=1}^n [i]_q\cdot \prod_{i=1}^n (q^i+1).
\end{equation*}
The Weyl group of $P(\flis F)$ is $W_J = A_{np_1-1}\times\cdots\times A_{np_{k-1}-1}\times C_{np_k}$, whence
\begin{equation*}
    P_{W_J}(q) = \prod_{i=1}^{k-1}\prod_{j=1}^{np_i}[j]_q\cdot \prod_{j=1}^{np_k}[j]_q\,(q^j+1),
\end{equation*}
and \eqref{eq:point_count} yields the exact count
\begin{equation}\label{eq:sp_flag_count}
    |\Sp(V)/P(\flis F_{n,P})| = \binom{n}{nP}_q \prod_{j=np_k+1}^{n}(q^j+1),
\end{equation}
where $\flis F_{n,P}$ denotes any isotropic flag of type $(np_1,\dots,np_{k-1})$. Appendix \ref{app:symplectic} verifies \eqref{eq:sp_flag_count} by elementary arguments---Witt decompositions and explicit matrix counts---independently of the general theory.

The following result is the special case of Proposition \ref{prop:lie_entropy} for the symplectic group; thanks to \eqref{eq:sp_flag_count}, it admits a direct and self-contained proof.

\begin{proposition}\label{prop:spentropy}
    Let $P=(p_1,...,p_k)$ be a probability vector with rational entries and  $N_P$ be the set of positive integers $n$ such that $np_i\in \Z$ for all  $i=1,...,k$. Let $V=\Fq^{2n}$ be a symplectic space and $\flis F_{n,P}$ be an isotropic flag of type $(np_1,np_2,\dots, np_{k-1})$. Then, \begin{equation*}\label{eq:spentropy}
  \lim_{\substack{ n\rightarrow\infty \\ n\in N_P}} \frac{1}{n^2} \log_q \left(|\Sp(V)/P(\flis F_{n,P})|\right) = \Hq(P) + \frac 1 2 (1-p_k^2).
\end{equation*}
\end{proposition}
\begin{proof}
By \eqref{eq:sp_flag_count},
 $$ |\Sp(V)/P(\flis F_{n,P})|  = \binom{n}{nP}_q \prod_{j=np_k+1}^n (q^j+1).$$
In the limit
\begin{align*}
    & \lim_{n\rightarrow\infty} \frac{1}{n^2} \log_q \left(\frac{|\Sp(V)|}{|P(\flis F_{n,P})|}\right) \\
    &= \lim_{n\rightarrow\infty} \big( \frac{1}{n^2} \log_q  \binom{n}{nP}_q \big) + \lim_{n\rightarrow\infty} \big( \frac{1}{n^2} \log_q \prod_{j=np_k+1}^n (q^j+1) \big) \\
    &= \lim_{n\rightarrow\infty} \big(\Hq(P) + o(1) \big) + \lim_{n\rightarrow\infty}  \big(\frac{1}{2}(1-p_k^2) + o(1) \big) \\
    &= \Hq(P) + \frac{1}{2}(1-p_k^2)
\end{align*}
\end{proof}

\begin{remark}[Typicality over $\Fq$]\label{rem:flag_typicality}
    Part of the typicality picture of Section \ref{sec:typicality} survives over $\Fq$, with the strata of flags of a given type playing the role of type classes. Fix the length $k$. The type of an isotropic flag records the dimensions $0<d_1<\cdots<d_{k-1}\le n$ of its members, so the number of types is at most $\binom{n+k-1}{k-1} = O(n^{k-1})$, i.e. \emph{polynomially many} in $n$ \cite[\S1.2]{stanleyEC1}. Each stratum, by contrast, is \emph{exponentially large}, of size $\exp_q(n^2\HSp + O(n))$. This is exactly the setting of the method of types \cite{csiszar1998}: exponentially many objects sorted into polynomially many classes that are separated on the scale $n^2\ln q$. Since the polynomial count of classes contributes only $O(\log n)$ to $\log_q$, the largest stratum dominates the sum---as in \eqref{eq:typical_set_rate}, the logarithm of the \emph{total} number of flags grows like that of the largest stratum---and the uniform measure concentrates on it. Concretely, take flags of length $k=2$, i.e. the set of \emph{all} isotropic subspaces $S\subset \Fq^{2n}$ (all dimensions combined) under the uniform measure. Writing $\dim S = \sigma n$, the stratum count \eqref{eq:sp_flag_count} is $|\IG(\sigma n,\Fq^{2n})| = \exp_q\!\big(n^2\,\HSp(\sigma,1-\sigma) + O(n)\big)$, with error uniform in $\sigma$ since it is an explicit $q$-product. The rate $\sigma\mapsto \HSp(\sigma,1-\sigma) = 2\sigma - \frac 32 \sigma^2$ is strictly concave with a unique interior maximizer $\sigma^* = 2/3$, so any $\sigma$ with $|\sigma-\tfrac23|\ge\varepsilon$ carries probability at most $(n+1)\exp_q\!\big({-}\delta(\varepsilon)\,n^2 + O(n)\big)\to 0$: the normalized dimension $\dim S/n$ concentrates around $\sigma^*=2/3$, and a uniformly random isotropic subspace of $\Fq^{2n}$ has dimension close to $2n/3$. This is the  analogue of the fact that a uniformly random word of length $n$ has approximately balanced type, and of the fact that a uniformly random subspace of $\Fq^{n}$ has dimension close to $n/2$, the maximizer of $\Hq(\sigma,1-\sigma)$; cf. \cite{vigneauxfinite}. At this stage, however, the correspondence holds only at the level of a single measure. Classically the uniform law on length-$n$ words is the time-$n$ marginal of a  \emph{memoryless source}---a consistent family of laws generated by a dynamics---whereas here we have only a \emph{sequence} of uniform measures on the flags of $\Fq^{2n}$, with no dynamics passing from $\Fq^{2n}$ to $\Fq^{2(n+1)}$ and no consistency relating them. The notion of flag-valued (or even subspace-valued) source remains mostly undeveloped, with the exception of the subspace-generating process introduced by Vigneaux \cite{vigneauxfinite} in connection with the $q$-binomial theorem (however, its marginals are not uniform).\footnote{This differs from two established uses of subspaces in information theory: subspace codes \cite{koetterkschischang}, where a subspace is a \emph{channel} codeword rather than the output of a source; and the \emph{typical subspace} of quantum data compression \cite{schumacher1995}, which is a subspace of the Hilbert space carrying an ordinary quantum source, not a subspace-valued random variable.} One can nonetheless enlarge the supply of \emph{one-shot} laws to reach non-uniform types, by tilting the uniform measure into the Gibbs ensemble of Remark \ref{rem:gibbs_flags} below; this still produces static measures, not a process. We leave the construction of flag-generating sources with prescribed marginals to future work.
\end{remark}

\begin{remark}[A Gibbs ensemble of flags]\label{rem:gibbs_flags}
    The uniform measure of Remark \ref{rem:flag_typicality} concentrates on the single type maximizing $\HSp$. To single out any other type by a \emph{one-shot} law, we can tilt the uniform measure by a linear function of the type. Write $P=(p_1,\dots,p_{k-1})$ for the free coordinates of a type (with $p_k = 1-\sum_{i<k}p_i$ determined), so that $\mathrm{type}(\flis F) = nP$ is the composition recording the successive quotient dimensions of $\flis F$, and, for $\theta\in\mathbb R^{k-1}$, set
    \begin{equation*}
        \mu_\theta(\flis F)\propto\exp_q\!\big(n\,\langle\theta,\mathrm{type}(\flis F)\rangle\big)=\exp_q\!\big(n^2\langle\theta,P(\flis F)\rangle\big).
    \end{equation*}
    The factor $n$ is essential: it places the tilt at the scale $n^2\ln q$ of the entropy, whereas a fixed $\theta$ paired with the $O(n)$ dimension vector would give a tilt $\exp_q(O(n)) = \exp_q(o(n^2))$, washed out by the error term and leaving the typical type unchanged.

    Since $\mu_\theta$ depends on $\flis F$ only through its type, and there are $\exp_q(n^2\HSp(P)+O(n))$ flags of type $nP$ by \eqref{eq:sp_flag_count}, the induced law on types is
    \begin{align*}
        \mu_\theta(\mathrm{type}=nP)&=\frac{\exp_q\!\big(n^2[\HSp(P)+\langle\theta,P\rangle]+O(n)\big)}{Z_n(\theta)},\\
         \tfrac1{n^2}\log_q Z_n(\theta)&\xrightarrow[n\to\infty]{}\Lambda(\theta),
    \end{align*}
    where $\Lambda(\theta):=\max_{P}\big[\HSp(P)+\langle\theta,P\rangle\big]$; the limit is  the maximum over the polynomially many types dominating the sum, as one can establish following Laplace's method. As $\HSp$ is strictly concave on the simplex (its Hessian is negative definite), the maximizer is unique and $\mu_\theta$ concentrates on it,
    \begin{equation*}
        P_\theta^\ast=\arg\max_P\big[\HSp(P)+\langle\theta,P\rangle\big],\qquad \nabla\HSp(P_\theta^\ast) = -\theta.
    \end{equation*}
    Conversely, any type $P$ in the interior of the simplex is the typical type of $\mu_\theta$ for $\theta = -\nabla\HSp(P)$: the family $\{\mu_\theta\}_{\theta\in\mathbb R^{k-1}}$ realizes \emph{arbitrary} interior types. This is a Gibbs ensemble with natural parameters $\theta$, free energy $\Lambda$, and entropy $\HSp$, the two being Legendre dual ($P_\theta^\ast = \nabla\Lambda(\theta)$); it is the finite-field, $n^2$-speed counterpart of the tilted (exponential) families that classically generate non-uniform types \cite[\S2.2]{dembozeitouni}.
\end{remark}

\begin{remark}[Stabilizer codes]\label{rem:stabilizer}
    The counts of this section have a quantum-information interpretation. In the stabilizer formalism over qudits of prime-power dimension $q$, a stabilizer group on $n$ qudits corresponds, modulo phases, to a totally isotropic subspace of the phase space $\Fq^{2n}$ equipped with its standard symplectic form, and an $[[n,k]]_q$ stabilizer code to an isotropic subspace of dimension $n-k$. The number of such codes, computed in \cite[Thm.~20]{gross2006} and \cite{singal2023}, is precisely $|\IG(n-k,\Fq^{2n})|$ of Proposition \ref{prop:IG}---the case of flags of length one in \eqref{eq:sp_flag_count}. The counting argument of \cite{singal2023} runs in parallel with Appendix \ref{app:symplectic}: their group $T(2n,k,\Z_d)$, called for $k=0$ the ``Borel subgroup'' in the literature on Bruhat decompositions of Clifford circuits (see \cite[Rem.~5]{singal2023}), is the parabolic subgroup $P(S)$, and their factorization of $C_{[[n,k]]}$ matches \eqref{eq:card_P}. Proposition \ref{prop:spentropy} therefore computes the growth rate of the number of stabilizer codes of rate $\kappa = k/n$:
    \begin{equation*}
        \#\big\{[[n,\kappa n]]_q \text{ stabilizer codes}\big\} = \exp_q\!\big(n^2 \HSp(1-\kappa,\kappa) + o(n^2)\big),
    \end{equation*}
    and, by Remark \ref{rem:flag_typicality}, a stabilizer code drawn uniformly at random (all $k$ combined) has rate close to $1/3$. Isotropic flags correspond to nested stabilizer groups, so the chain rule below admits a coding interpretation as well. (For qudits of non-prime-power dimension $d$ the relevant group is the symplectic group over the ring $\Z_d$, and the growth rate changes; see \cite{singal2023}.)
\end{remark}

\subsection{Chain rule}\label{sec:asymptotics_symplectic}

In parallel with Proposition \ref{prop:chain_rule_reflective}, the chain rule of the Lie entropy is an asymptotic shadow of the Poincaré-polynomial identity \eqref{eq:coarseningdynkin}---this time evaluated at $t=q$, where by \eqref{eq:point_count} it becomes an identity between cardinalities of flag varieties over $\Fq$.

\begin{proposition}[Chain rule]\label{prop:chain_rule_sp}
    Let $\pi:\{1,\dots, k\}\to \{1,\dots, m\}$ be an increasing surjection. Let $P$ be a probability vector over $\{1,\dots, k\}$, and let $Q$ denote the push-forward measure $\pi_*P$ given by $q_j = \sum_{i\in \pi\inv(j)}p_i$ for $j=1,\dots, m$. Set $P|_{\pi^{-1}(j)} \coloneqq (\frac{p_i}{q_j})_{i\in\pi^{-1}(j)}$.   The Lie entropies $ \HSp(P)$ and $\HSp(Q)$ are related by the formula
    \begin{equation*}\label{eq:sympchain1}
       \HSp(P) = \\ \HSp(Q) +  \sum_{j=1}^{m-1} q_j^2\, \Hq(P|_{\pi^{-1}(j)}) + q_m^2 \HSp(P|_{\pi^{-1}(m)}).
    \end{equation*}
\end{proposition}

\begin{proof}
    Let $W$ be of type $B_n$, $C_n$ or $D_n$---so that $G$ is $\SO_{2n+1}$, $\Sp_{2n}$ or $\SO^+_{2n}$, respectively---with fundamental root system $\Delta = \{1,\dots,n\}$ and $n\in N_P$, and write $X\in\{B,C,D\}$ for its type. Let $I\subset J\subset \Delta$ be the complements of the sets of nodes $\{np_1, n(p_1{+}p_2),\dots, n(1{-}p_k)\}$ and $\{nq_1, n(q_1{+}q_2),\dots, n(1{-}q_m)\}$, respectively. The connected components in $\mathbb J$ are blocks of type $A_{nq_j-1}$, for $j=1,\dots,m-1$, and a tail $S^*$ of type $X_{nq_m}$: the component containing the distinguished end node inherits the type of $W$, while the interior blocks are of type $A$ regardless.

    We evaluate \eqref{eq:coarseningdynkin} at $t=q$ and interpret each factor via \eqref{eq:point_count}: the quotients $P_W(q)/P_{W_I}(q)$ and $P_W(q)/P_{W_J}(q)$ count the totally isotropic flags of types $nP$ and $nQ$ for the form defining $G$ (in the symplectic case, flags in $\Fq^{2n}$, cf. \eqref{eq:sp_flag_count}; in general, cf. Proposition \ref{prop:lie_entropy}); for an $A$-type component $S\in\mathbb J\setminus\mathbb I$ corresponding to $j<m$, the quotient $P_{W_S}(q)/P_{W_{S\cap I}}(q) = \binom{nq_j}{nq_j P|_{\pi^{-1}(j)}}_q$ counts ordinary flags in $\Fq^{nq_j}$; and for the tail, $P_{W_{S^*}}(q)/P_{W_{S^*\cap I}}(q)$ counts totally isotropic flags of type $nq_m\, P|_{\pi^{-1}(m)}$ for a form of the same type $X$ in rank $nq_m$.

    Now apply $\frac{1}{n^2}\log_q$ and take the limit $n\to\infty$ along $N_P$. By Proposition \ref{prop:lie_entropy}---which yields the \emph{same} Lie entropy $\HSp$ for all three types $B$, $C$, $D$---the isotropic counts converge to $\HSp(P)$ and $\HSp(Q)$, and, replacing $n$ by $nq_m$, the tail of type $X_{nq_m}$ contributes $q_m^2\, \HSp(P|_{\pi^{-1}(m)})$ irrespective of $X$; by \eqref{eq:limit-GL}, the $A$-type factor of index $j$ contributes $q_j^2\, \Hq(P|_{\pi^{-1}(j)})$. Components belonging to both $\mathbb I$ and $\mathbb J$ correspond to Dirac conditional laws and contribute no entropy. In the symplectic case, the self-contained Proposition \ref{prop:spentropy} may replace Proposition \ref{prop:lie_entropy}.
\end{proof}

In the symplectic case ($C_n$), this identity has the following combinatorial interpretation: to build an isotropic flag $\flis F_{n,P}=(S_1 \subset \cdots S_{k-1})$ we can first select an isotropic flag $\flis G_{n,Q} = (S_1' \subset \cdots \subset S_{\ell -1})$, with induced Witt decomposition of the ambient symplectic space $V$ into $S'_{\ell-1} \oplus T'_{\ell-1} \oplus W' _{\ell-1}$ (see Appendix \ref{app:symplectic}), and then choose ordinary flags within $S_1'$ and each $S'_{i}/S'_{i-1}$ for $i=2,...,\ell-1$ (the spaces involved in these ordinary flags are automatically isotropic), while also choosing an isotropic flag in $W'_{\ell-1}$, which has dimension $2nq_m$, each of whose spaces can be used to extend the isotropic subspace $S'_{\ell-1}$ via direct sum.

\section{Discussion: dimensions, volumes, and continuous analogues}\label{sec:discussion}
We close with some remarks on the nature of the quantities computed in this article and their continuous counterparts.

First, the Lie-type entropies are, at bottom, \emph{dimension growth rates}. The proof of Proposition \ref{prop:lie_entropy} shows that
\begin{equation*}
    \frac 1{n^2}\log_q |(G_n/P_J)(\Fq)| = \frac{\dim (G_n/P_J)}{n^2} + O\left(\frac{\log n}{n}\right),
\end{equation*}
since the degree $D_n$ of the polynomial \eqref{eq:point_count} is the dimension of the flag variety (the top Schubert cell). In particular the limit is \emph{independent of $q$}: the Lie entropy is the asymptotic normalized dimension of the flag varieties of the series $B$, $C$, $D$, just as $\Hq$ is for the series $A$.

Second, counting points is, in a precise sense, computing a volume. For a smooth projective variety $X$ with good reduction, Weil's theory of canonical (Tamagawa) measures gives
\begin{equation*}
    \frac{|X(\F_q)|}{q^{\dim X}} = \int_{X(\Z_q)} |\omega|,
\end{equation*}
the integral of a canonical top form over the $q$-adic points \cite{weiladeles}. The cardinalities studied in Section \ref{sec:lie} are therefore integrals of top forms, i.e. volumes, over nonarchimedean points, with $\log q$ playing the role of the unit of dimension.

This suggests a bridge to the \emph{differential} entropy $h(f) = -\int f \ln f$ of a probability density on $\mathbb R$: for an absolutely continuous probability measure $P$ on $\mathbb R^d$ with marginal density $f$, the typical set $T^n_{[f]\delta}\subset \mathbb (R^d)^n$ carries most of the probability and has \emph{$nd$-volume} $\exp(n(h(f)+o(1)))$ \cite[Ch.~8]{coverthomas}; in other words
$$ \frac 1n \log (\mathrm{vol}(\text{typical set})) = h(f) + o(1).$$
Typically, the quantity $h(f)$ depends on the ambient dimension $d$, for instance, the differential entropy of a Gaussian law $N_d(\mu, \Sigma)$ with mean $\mu\in \mathbb R^d$ and covariance matrix $\Sigma$ is $\frac{d}{2} \log (2\pi \mathrm{e}) + \frac{1}{2} \log \det\Sigma$.

That differential entropy depends on the choice of unit (it is not invariant under rescaling) matches the fact that a volume, unlike a cardinality, requires a normalization; over $\Fq$ the normalization is canonical ($\log q$), which is why the limits of Proposition \ref{prop:lie_entropy} are $q$-independent. It is natural to ask for genuinely continuous analogues of our results---for instance, ensembles of flags in $\mathbb R^n$ or $\mathbb C^n$ (say, kernels or spans of random Gaussian matrices, or invariant measures on real and complex flag manifolds) whose typical sets have volume growth governed by functionals of the type $\HSp$. We leave this as an open direction.

More generally, given a probability measure $\rho$ on $\mathbb R^d$, quantized on the cubic mesh of side $1/n$, R\'enyi showed that (under suitable hypotheses) the entropy of the discretized law expands as
\begin{equation*}
    H(\rho_n) = a \ln n + b + o(1),
\end{equation*}
and called $a$ the \emph{information dimension} and $b$ the \emph{$a$-dimensional entropy} of $\rho$: for a discrete law, $a = 0$ and $b$ is the Shannon entropy; for an absolutely continuous one on $\mathbb R^d$, $a=d$ and $b$ is the differential entropy. We see some analogies with our results: the coefficient of $n^2$ in $\log_q|(G_n/P_J)(\Fq)|$ is the normalized dimension (which converges to the Lie entropy when $P_J$ is probabilistically defined) while finer invariants are carried by the lower-order coefficients; for instance, the parameter $e$ that distinguishes the polar spaces appears in the coefficient of $n$ (Remark \ref{rem:polar}). Graded expansions of this kind occur in type $A$ as well: Stirling's formula refines \eqref{eq:limit_binomial} into $\ln\binom{n}{nP} = nH(P) - \frac{k-1}{2}\ln n + O(1)$, where the coefficient of $\ln n$ is (half) the dimension of the simplex of types, hence it captures the number of symbols in the alphabet. A systematic generalization of Renyi's definitions to flags as well as a study of the subleading ``entropies'' seems worthwhile.

Finally, the two regimes studied in this article are tied together by a classical observation of Tits \cite{tits1957}: the coset space $W/W_J$ behaves in many respects as the flag variety of a ``field with one element'', the $q\to 1$ limit of $(G/P_J)(\Fq)$---an analogy made quantitative by \eqref{eq:point_count}, whose value at $q=1$ is precisely $|W/W_J|$. The reflective and Lie entropies are the asymptotic shadows of the two ends of this deformation, with their distinct normalizations ($n$ versus $n^2$) and limiting functionals ($H_R$ versus $\HSp$).

\section*{Acknowledgment}

The authors are grateful to Daniel Juteau for several enlightening remarks that had a deep impact in the final structure of this article. In particular, he pointed out the general product formula for the Poincaré polynomial of a flag variety $G/P$ in terms of the degrees of the Weyl group, which underlies the uniform treatment adopted in this article, as well as the coincidence of the Coxeter groups of types $B_n$ and $C_n$ (Remark \ref{rem:BC}). R.L. would like to thank Mr. and Mrs. Robert C. Loschke for their generous financial contribution to his Summer Undergraduate Research Fellowship, and Ms. Carol Casey for her patience and assistance throughout the process. All authors want to thank Prof. Matilde Marcolli for her encouragement and support. Claude Opus 4.8 and Fable 5 assisted with literature review, proofreading, and with drafting some mathematical arguments in Sections \ref{sec:signed_typicality} and \ref{sec:lie}; that said, all claims and proofs have been carefully verified by the authors. 

\appendix

\section{Proof of Factorization of Poincar\'e Polynomial}\label{app:factorpoincare}

Let $W$ be a reflection group with fundamental root system $\Delta$ with $\Delta = \Delta_X\cup\Delta_Y$ and $\Delta_X\cap\Delta_Y = \varnothing$ such that for any $\alpha\in W_X$ and $\beta\in W_Y$ we have that $g_\alpha, g_\beta\in W$ commute (in particular, $(g_\alpha g_\beta)^2 =1 $). Then we have that every element $g\in W$ can be factored as $g_x g_y$ for some $g_x\in W_X$ and $g_y\in W_Y$. Then we see that for any $\alpha\in \Delta_X$ and $\beta\in \Delta_Y$, we have that if $\alpha$ sends a fundamental root $\gamma$ to a negative root, then $\beta$ must not change the sign of that root, and vice versa. Thus \[\ell(g_\alpha g_\beta) = \ell(g_\alpha) + \ell(g_\beta)\]
So now we consider the Poincar\'e polynomial of $W=W_XW_Y$:
\begin{align*}
    P_{W_XW_Y}(t) &= \sum_{g\in W} t^{\ell (g)}
    =\sum_{g_1 g_2 \in W} t^{\ell(g_1g_2)} \\
    &=\sum_{g_1g_2\in W}t^{\ell(g_1)+\ell(g_2)}
    =\sum_{g_1g_2\in W} t^{\ell(g_1)} t^{\ell(g_2)} \\
    &= \left(\sum_{g_1\in W_X} t^{\ell (g_1)}\right)\left(\sum_{g_2\in W_Y} t^{\ell (g_2)}\right) \\
    &= P_{W_X}(t) P_{W_Y}(t)
\end{align*}

\section{The symplectic group: elementary counting}\label{app:symplectic}

This appendix verifies, by elementary arguments, the facts used in Section \ref{sec:symplectic}: the transitivity of the action of $\Sp(V)$ on the isotropic flags of a given type, and the exact count \eqref{eq:sp_flag_count}. Throughout, $(V,f)$ is a symplectic space of dimension $2n$ over $\Fq$, and for a subspace $U\subset V$ we write $U^\perp = \{v\in V\mid f(v,u)=0 \text{ for all } u\in U\}$.

\subsection{Witt decomposition and transitivity}

Let $S$ be a totally isotropic subspace, $\dim S =: s \leq n$. There is a decomposition of $V$ (Witt's decomposition),
\begin{equation}\label{eq:decomposition}
    V = S \oplus W \oplus T,
\end{equation}
such that $T$ is an isotropic subspace with $\dim T = s$, $W$ is a symplectic subspace of dimension $2(n-s)$, and the symplectic form $f$ restricts to a perfect pairing $f|_{S\times T}:S \times T \rightarrow \Fq$. Given \eqref{eq:decomposition}, we can write
\begin{equation}\label{eq:sympl_form_decomposed}
    f((s_1,t_1,w_1), (s_2,t_2,w_2)) = t_2^t s_1 - t_1^t s_2 + f(w_1,w_2).
\end{equation}
Moreover, $S^\perp = S \oplus W$, so the quotient $S^{\perp}/S$ is a symplectic space naturally isomorphic to $W$.

Given two isotropic subspaces $S,S'$ of the same dimension, there is a transformation $g\in \Sp(V)$ such that $g(S) = S'$: it is enough to take any invertible linear map $L:S\to S'$, and act on $T$ by $(L^{t})^{-1}$ to preserve the symplectic form. Hence $\Sp(V)$ acts transitively on $\IG(s,V)$, the set of isotropic subspaces of dimension $s$. The same argument applied to a basis of $S_{k-1}$ adapted to a flag $\flis F = (S_1\subset\cdots\subset S_{k-1})$ (the first $\dim S_i$ vectors spanning $S_i$) shows that $\Sp(V)$ acts transitively on the isotropic flags of a given type $(m_1,\dots,m_{k-1})$; in particular, flags of the same type have conjugate stabilizers, and $|\Sp(V)/P(\flis F)|$ is the number of isotropic flags of that type.

\subsection{Counting isotropic subspaces}

Let $P(S)$ be the stabilizer of $S\in \IG(s,V)$ and consider the map
\[
    \phi: P(S) \rightarrow \GL(S) \times \Sp(W), \, g\mapsto (g_{|S}, \overline{g}),
\]
where $g_{|S}$ is the restriction of $g$ to $S$ and $\overline{g}$ is the map induced on $S^{\perp}/S \cong W$.

\begin{lemma}\label{lem:phi}
    $\phi$ is a surjective group homomorphism.
\end{lemma}
\begin{proof}
    For $g,h \in P(S)$,
    \begin{equation*}
        \phi(gh) = ((gh)_{|S}, \overline{gh}) = (g_{|S} \circ h_{|S}, \overline{g} \circ \overline{h}) = \phi(g)\phi(h),
    \end{equation*}
    where the second equality holds because $h(S)=S$. For surjectivity, let $(A,B) \in \GL(S) \times \Sp(W)$. Because the pairing $f|_{S\times T}$ is perfect, there is a unique linear map $A^*: T \rightarrow T$ such that $f( As, A^*t ) = f( s, t )$ for all $s \in S$, $t \in T$ (that is, $A^*$ is the adjoint of $A^{-1}$ with respect to the pairing). Define $g(s+w+t) \coloneqq As+Bw+A^*t$; by \eqref{eq:sympl_form_decomposed}, $g$ preserves $f$, so $g \in \Sp(V)$, $g(S)=S$ and $\phi(g) = (A,B)$.
\end{proof}

The kernel $N(S) \coloneqq \ker \phi$ thus fits in a short exact sequence $1 \rightarrow N(S) \rightarrow P(S) \xrightarrow{\phi} \GL(S) \times \Sp (W) \rightarrow 1$, so that
\begin{equation}\label{eq:card_P}
    |P(S)| = |N(S)| \cdot |\GL(S)| \cdot |\Sp (W)|.
\end{equation}
The three factors are classical counts. It is well-known that\footnote{Choosing an invertible matrix is equivalent to choosing first any nonzero vector (of which there are $q^m-1$), then any vector outside the one-dimensional space generated by the first one (of which here are $q^m-q$), and so on.} $|\GL_m(\F_q)|= \prod_{i=0}^{m-1} (q^m-q^i)$, and $|\Sp_{2n}(\Fq)| = q^{n^2} \prod_{i=1}^n (q^{2i}-1)$ \cite[Thm. 3.12]{classicalgrove}. As for the kernel, following \cite{genevievehanlon, 18704mitsupp},
\begin{equation*}
     N(S) = \{ N_AZ_E \mid A: T\to W \text{ linear; }E \text{ a }s\times s\text{ symmetric matrix}\},
\end{equation*}
where $Z_E(s,t,w) = (s+Et,t,w)$ and $N_A(s,t,w) = (s + C(w) + Q(t), t, w+A(t))$, with $C$ and $Q$ linear corrections uniquely determined by $A$ that make $N_A$ symplectic \cite{18704mitsupp}; a direct computation yields
\begin{equation}\label{eq:card_N}
    |N(S)| = q^{s(s+1)/2 + 2s(n-s)}.
\end{equation}
By transitivity, $|\IG(s,V)| = |\Sp(V)|/|P(S)|$; replacing the four cardinalities and simplifying \cite{genevievehanlon}, one obtains:

\begin{proposition}\label{prop:IG}
Let $V$ be a symplectic space of dimension $2n$ and $s$ an integer in $[0,n]$. Then,
    \begin{equation}
        |\IG(s,V)| = \frac{[n]_q!}{[s]_q![n-s]_q!} \prod_{j=n-s+1}^n (q^j+1)
    \end{equation}
\end{proposition}

\subsection{Counting isotropic flags}

\begin{proposition}\label{prop:counting_flags} Let $V$ be a symplectic vector space and $\flis F= (S_1 \subset \cdots S_{k-1})$ an isotropic flag of type $(m_1,...,m_{k-1})$. Set $s=\dim S_{k-1} = \sum_{i=1}^{k-1}m_i$. Then
    $$|\Sp(V)/P(\flis F)| = |\IG(s,V)| | F(m_1,...,m_{k-1}; \F_q)| $$
    where $F(m_1,...,m_{k-1}; \F_q)$  is the space of ordinary flags of type $(m_1,...,m_{k-1})$ inside $S_{k-1}$, counted by $${s \choose m_1, \dots m_{k-1}}_q$$
\end{proposition}
\begin{proof}
    We simply remark that to build a flag $\flis F=(S_1 \subset \cdots S_{k-1})$, it is enough to pick a totally isotropic subspace (of which there are $|\IG(s,V)|$) and then \emph{any} flag of subspaces $(S_1\subset \cdots S_{k-2} \subset S_{k-1})$, since these will automatically be isotropic because $f|_{S\times S} \equiv 0$.
\end{proof}

Finally, let the flag $\flis F$ have type $(np_1,\dots,np_{k-1})$ for a probability vector $P=(p_1,\dots,p_k)$, so that $s = (1-p_k)n$. Since
$$|\IG((1-p_k)n,V)| =\frac{[n]_q!}{[p_kn]_q![(1-p_k)n]_q!} \prod_{j=p_kn+1}^n (q^j+1)$$
by Proposition \ref{prop:IG}, and
$$ | F(p_1n,...,p_{k-1}n; \F_q)| =  \frac{[(1-p_k)n]_q!}{\prod_{i=1}^{k-1} [np_i]_q!}, $$
Proposition \ref{prop:counting_flags} reproduces the exact count \eqref{eq:sp_flag_count}, in agreement with the general product formula \eqref{eq:point_count}.

%%%%%%
%% To balance the columns at the last page of the paper use this
%% command:
%%
%\enlargethispage{-1.2cm} 
%%
%% If the balancing should occur in the middle of the references, use
%% the following trigger:
%%
%\IEEEtriggeratref{4}
%%
%% which triggers a \newpage (i.e., new column) just before the given
%% reference number. Note that you need to adapt this if you modify
%% the paper.  The "triggered" command can be changed if desired:
%%
%\IEEEtriggercmd{\enlargethispage{-20cm}}
%%
%%%%%%

%%%%%%
%% References:
%% We recommend the usage of BibTeX:
%%
\bibliographystyle{abbrvnat}
%\bibliography{definitions,bibliofile}
\bibliography{sample-base}
%%
%% where we here have assumed the existence of the files
%% definitions.bib and bibliofile.bib.
%% BibTeX documentation can be obtained at:
%% http://www.ctan.org/tex-archive/biblio/bibtex/contrib/doc/
%%%%%%

%% Or you use manual references (pay attention to consistency and the
%% formatting style!):

\end{document}